\documentstyle[12pt,epsf]{article}

\begin{document}

\title{The Static Quark-Antiquark Potential in QCD to Three Loops}
\author{Markus Peter \\
 {\normalsize Institut f\"ur Theoretische Teilchenphysik,}
 \\{\normalsize  Universit\"at Karlsruhe} \\ {\normalsize
  D-76128 Karlsruhe, Germany}}
\date{October 1, 1996}
\maketitle
\thispagestyle{empty}
\vspace{-4.0truein}
\begin{flushright}
{\bf TTP 96-37\footnote{
    The complete paper, including figures, is
    also available via anonymous ftp at
    ttpux2.physik.uni-karlsruhe.de (129.13.102.139) as
    /ttp96-37/ttp96-37.ps,
    or via www at
    http://www-ttp.physik.uni-karlsruhe.de/cgi-bin/preprints}
}\\
{\bf October 1996}\\
{\bf hep-ph/9610209}
\end{flushright}
\vspace{3.0truein}
\begin{abstract}
  The static potential between an infinitely heavy quark and antiquark
  is derived in the framework of perturbative QCD to three loops by
  performing a full calculation of the two-loop diagrams and using the
  renormalization group. The contribution of massless fermions is
  included.
\end{abstract}


The force law between infinitely heavy quarks has been investigated since more
than 20 years because of its importance for a deeper understanding of the
strong interactions. The static quark-antiquark potential is a very
fundamental concept, constituting the non-abelian analog to the Coulomb
potential of electrodynamics, and also enters as a vital ingredient in the
description of non-relativistic bound states like quarkonia.  It is widely
believed to consist of two parts: a Coulombic term at short distances which
can be derived from field theory by using perturbative QCD, and a long-ranged
confining term whose derivation from first principles presumably requires much
more advanced methods. Although an analysis based on perturbation theory alone
thus cannot give the complete potential, the result of such an effort would
nevertheless be very useful. It could provide an improved input for QCD
inspired potential models or even describe very heavy systems to a reasonable
accuracy by itself. It could be compared with the potential obtained from
numerical studies using lattice gauge theory and it might also give some hints
on the nonperturbative regime.

The first investigation of the static perturbative QCD potential has been
performed in \cite{Susskind}. Although this work has been extended by several
groups shortly after \cite{F77,ADM77,ADM78,Bil80}, and some of these also
studied aspects of the two- \cite{F77} and even three-loop diagrams
\cite{ADM78}, there is still no full calculation of the two-loop diagrams
available. The purpose of the present paper is to fill this gap and hence,
by exploiting the renormalization group equation,
to obtain the three-loop potential.

Before turning to the actual analysis, let us first recall the
calculational procedure employed. It seems appropriate to begin with the
simplest case, the Abelian theory without massless fermions.

The static potential in QED can be defined in a way which makes its gauge
invariance manifest via the vacuum expectation value of a Wilson loop taken
about a rectangle of width $R$ and length $T\gg R$:
\begin{equation}
  V(R) = -\lim_{T\to\infty}\frac{1}{iT}\ln
        \langle P\exp\Big( ie \oint dx_\mu A^\mu \Big) \rangle
\end{equation}
where $P$ denotes the path ordering prescription.

The functional integral can be calculated exactly, and one indeed finds the
Coulomb potential plus an additional term which represents the self-energy of
the sources \cite{F77}. To compare with the non-Abelian theory it is, however,
useful to go through the perturbative analysis as well. The Feynman rules for
the source are as follows: a source-photon vertex corresponds to a factor
$iev^\mu$, with an additional minus sign for the ``anti-source'', and the
source ``propagator'' reads
\begin{equation} \label{propx}
  S_F(x-x^\prime) = -i\Theta(x_0-x_0^\prime)\delta({\bf x}-{\bf x}^\prime)
\end{equation}
in coordinate space or
\begin{equation} \label{propp}
  S_F(p) = \frac{1}{vp+i\varepsilon}
\end{equation}
in momentum space. The four vector $v$ is given by $ v^\mu = (1,{\bf 0}) $ and
has only been introduced for notational reasons. The appearance of a
propagator for the sources is a consequence of the time ordering prescription
in the path integral, which introduces $\Theta$-functions when expanding the
exponential.

Looking at the problem in this way shows the connection to another approach:
the potential beyond the infinite mass limit or for non-singlet sources is
frequently also derived from the scattering operator; see for example
\cite{GR81,LS89}. The static QED potential thus should be derivable from the
scattering operator of heavy electron effective theory, the QED analog of
heavy quark effective theory, and the Feynman rules given above obviously
support this point of view.

Some care is required because the Feynman diagrams do not directly correspond
to the potential but to $\exp[-iV(R)T]$. The consequence is that in the
abelian theory the one-photon exchange amplitude already gives the final
result:
$$ 
   -ie^2 \int\!dx_0dy_0\;v_\mu v_\nu D^{\mu\nu}(x-y) = 
   -ie^2T\int\!\frac{d^3q}{(2\pi)^3}\frac{e^{i{\bf qR}}}{-{\bf q}^2}
$$
where $D^{\mu\nu}$ is used to represent the photon propagator and ${\bf x}=0$,
${\bf y}={\bf R}$ are understood.

\begin{figure}
  \begin{center} \epsfxsize10cm \mbox{\epsfbox[40 680 245 750]{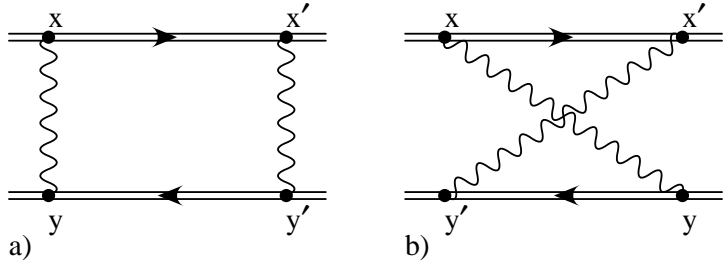}}
  \end{center}
  \caption{One-loop ladder diagrams. The double lines indicate static sources.}
  \label{oneloop}
\end{figure}

At one loop order, one encounters self-energy and vertex corrections which
cancel due to the Ward identity, and the ladder diagrams shown in Fig.\ 
\ref{oneloop} which are best analyzed in coordinate space. Because of the
simple structure of the source propagator, Eq.\ (\ref{propx}), only
integrations over time-variables remain. Adding the two diagrams removes the
$\Theta$-function corresponding to the anti-source propagator, and adding them
once more with $x\leftrightarrow x^\prime$, the source propagator can also be
removed and the one-photon exchange squared is obtained,
$$
 2\times\mbox{Fig.}~\ref{oneloop} 
  = \Big(-ie^2T \int\!dx_0dy_0\;D^{00}(x-y)\Big)^2.
$$
This behaviour of the ladder diagrams persists in higher orders \cite{F77},
the exponential thus starts to build up.

To see the exponentiation in momentum space is more difficult, and requires
that we specify the gauge --- which will be Feynman gauge --- and the special
kinematic situation. As the sources are infinitely heavy they may carry any
three-momenta without moving, but the actual values of these three-momenta are
irrelevant as the only quantities that enter the calculation are the momentum
{\em transfer} $q$ and the energies of the sources. The latter are required to
vanish by the on-shell condition (implied in the propagator, for example) and
consequently the only dimensionful parameter that remains is $q^2=-{\bf q}^2$.
Using dimensional regularization with $D=4-2\epsilon$ to handle infrared
divergencies, the individual amplitudes for the diagrams in Fig.\ 
\ref{oneloop} thus read
\begin{eqnarray}
  \mbox{Fig.}~\ref{oneloop}(a) & = & ie^2\Big(\frac{e}{4\pi}\Big)^2
    \frac{4}{q^2} \Gamma(\epsilon)\Big(\frac{4\pi\mu^2}{-q^2}\Big)^\epsilon
    \;r(v\cdot v^\prime), \\
  \mbox{Fig.}~\ref{oneloop}(b) & = & ie^2\Big(\frac{e}{4\pi}\Big)^2
    \frac{4}{q^2} \Gamma(\epsilon)\Big(\frac{4\pi\mu^2}{-q^2}\Big)^\epsilon
    \;r(-v\cdot v^\prime),
\end{eqnarray} 
with
\begin{eqnarray}
  r(w) & = & \frac{1}{\sqrt{w^2-1}}\ln\Big(w+\sqrt{w^2-1}\Big) \qquad,~w>1, \\
  r(w) & = & -r(-w) +\frac{i\pi}{\sqrt{w^2-1}} \qquad,~w<-1.
\end{eqnarray}
$v^\prime$ is used to denote the four-velocity of the anti-source, hence we
should take the limit $v^\prime\to-v$, $v\cdot v^\prime \to -1$. But this
results in a badly diverging imaginary part for the uncrossed diagram. Keeping
a relative motion by setting $v^\prime = (-\sqrt{1+{\bf u}^2},{\bf u})$ we
can, however, recognize this divergence as resulting from the Coulomb phase
\cite{KF71}. Another way to see this fact would be to keep the kinetic energy
in the heavy electron propagator. As the real parts of the diagrams cancel we
thus again find that they are merely an iteration of the one-photon exchange.
A similar analysis should be possible for the higher order ladder diagrams of
course, but it is obvious that the coordinate space approach is much easier in
this respect.

The inclusion of $n_f$ massless (i.e., $m^2\ll{\bf q}^2$) fermions, although
it makes an exact solution impossible, presents no problem in perturbation
theory. The fermions appear as loops in the photon propagator and induce
light-by-light scattering and in this way lead to an effective running
coupling constant, i.e.,
\begin{equation}\label{alphav}
   V({\bf q}^2) = -\frac{4\pi\alpha_{\rm V}({\bf q}^2)}{{\bf q}^2}.
\end{equation}
Note that this effective coupling differs from the usual running coupling in
the $\overline{\rm MS}$-scheme. Light-by-light scattering, in fact, first
enters in three-loop graphs and is thus beyond the scope of this paper.

When turning to the non-Abelian case, the Wilson loop must be generalized to
$$
  \langle \mbox{Tr}\;P\exp\Big( ig \oint dx_\mu A^\mu_a T^a \Big) \rangle
$$
where the matrices $T^a$ denote the group generators. Consequently the
potential for a quark-antiquark pair in a color-singlet state can be defined
as
\begin{equation}
  V(R) = -\lim_{T\to\infty}\frac{1}{iT}\ln
    \langle \mbox{Tr}\;P\exp\Big( ig \oint\!dx_\mu\;A^\mu_aT^a \Big) \rangle.
\end{equation}
In principle there are some problems connected to this definition, caused by
the non-trivial topological structure of non-Abelian theories, which are,
however, absent in the purely perturbative approach.

As there is no way known to solve the QCD functional integral exactly, one has
to resort to a perturbative treatment, which is, of course, more complicated
than in the Abelian case: additional diagrams appear due to the trilinear and
quartic gluon self couplings, and the presence of the generators in the
source-gluon vertex influences the exponentiation as will be demonstrated.

We will use Feynman gauge and the kinematics as described above again.
Because the individual loop diagrams contain both infrared and ultraviolet
divergencies, dimensional regularization will be employed, without, however,
explicitly distinguishing between the two kinds of divergencies. The
$\overline{\rm MS}$-scheme will be adopted for renormalization.

The only difference between the non-Abelian and the Abelian theory on tree
level is the color factor $C_F= T_F(N^2-1)/N$ which multiplies the coupling
constant in the potential, where $N$ is the number of colors and $T_F$ the
normalization of the generators, $\mbox{Tr}[T^aT^b]=\delta^{ab}T_F$. It is
thus convenient to define
\begin{equation}
  V({\bf q}^2) = -C_F\frac{4\pi\alpha_{\rm V}({\bf q}^2)}{{\bf q}^2},
\end{equation}
as this allows for an immediate generalization to sources in the adjoint
representation: replacing $T^a_{kl}\to -if^{akl}$ and hence $C_F\to C_A=N$,
the function $\alpha_{\rm V}$ describes the potential for static gluinos as
well.

On the one-loop level the difference between QED and QCD is more prominent.
An obvious point is that the trilinear gluon self-coupling leads to a
correction to the gluon propagator even if $n_f=0$, and in principle to an
additional vertex correction as well. But as a consequence
of Feynman gauge and the special kinematics, every diagram containing a three
gluon vertex with all three ends directly attached to the sources vanishes: if
we denote the three-momenta flowing into the vertex with $p,q,r$, such a
diagram involves
$$
  v_\mu v_\nu v_\rho\Big( g^{\mu\nu}(p-q)^\rho + g^{\nu\rho}(q-r)^\mu
    +g^{\rho\mu}(r-p)^\nu\Big) = 0.
$$
The same statement holds for the four-gluon vertex, which, however, first
enters at the two-loop level.

A second and more interesting point is that the color factors associated
with the individual diagrams are not the same. Consider, for example, the
ladder diagrams of Fig.\ \ref{oneloop} again,
$$
  \mbox{Fig.}~\ref{oneloop}(a) \propto C_F^2\quad,\quad
  \mbox{Fig.}~\ref{oneloop}(b) \propto C_F^2-C_F\frac{C_A}{2}.
$$
We can immediately identify the terms $\propto C_F^2$ as iterations of the
tree-level potential, but there remains a term $\propto C_FC_A$ from the
crossed ladder which, together with a corresponding term from the vertex
correction that renders it infrared finite, leads to an additional
contribution to the one-loop potential. This, of course, influences the way the
exponentiation works at the two-loop level.

\begin{figure}[ht]
  \begin{center} \epsfxsize12cm \mbox{\epsfbox[45 550 560 760]{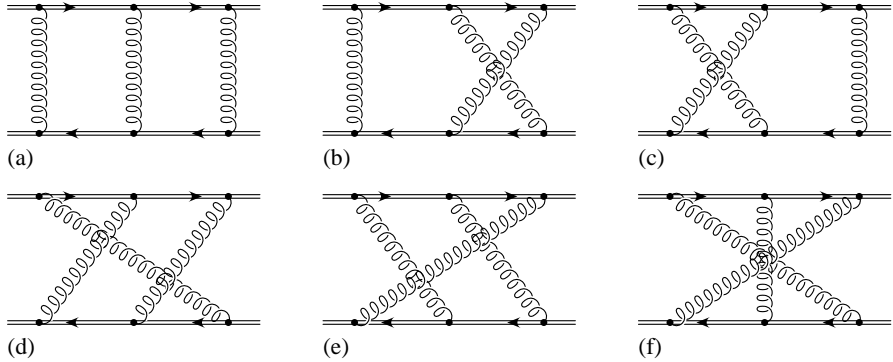}}
  \end{center}
  \caption{Two-loop ladder diagrams.}
  \label{twoloop}
\end{figure}

As has been demonstrated in \cite{F77}, the consequence is that in order to
compute the actual two-loop contribution to the potential, only those diagrams
have to be considered which involve color factors different from $C_F^3$ and
$C_F^2C_A$ and thus cannot result from iterations of the lower order
diagrams. This means that for example the first three of the ladder diagrams
in Fig.\ \ref{twoloop} and all graphs which are merely source self-energy
insertions in one-loop graphs are irrelevant.

\begin{figure}[ht]
  \begin{center} \epsfxsize12cm \mbox{\epsfbox[45 550 560 780]{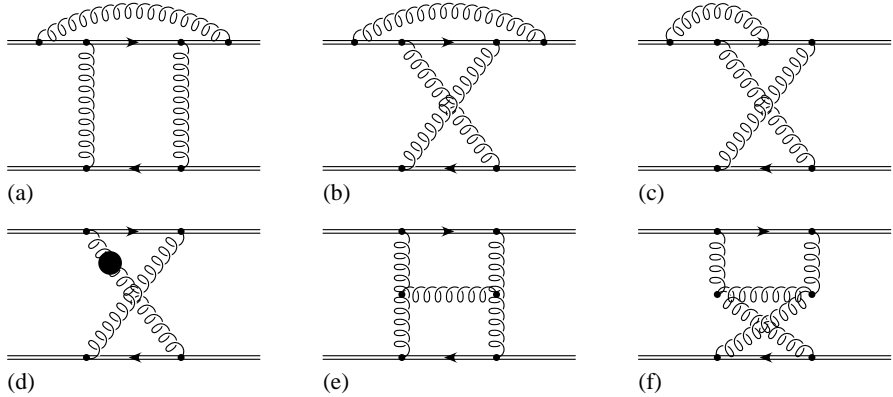}}
  \end{center}
  \caption{Corrections to one-loop ladder diagrams required for the potential
  calculation.}
  \label{mtl}
\end{figure}

To be more specific, the following diagrams have to be calculated:
\begin{itemize}
\item The two-loop ladder diagrams Fig.\ \ref{twoloop}(d-f).
\item The corrections to the one-loop ladder diagrams shown in Fig.\ 
  \ref{mtl}.  In general there would be more graphs of this type containing
  the three gluon vertex, e.g., analogous to Fig.\ \ref{mtl}(c), which,
  however, vanish in Feynman gauge as already explained. In fact, Fig.\ 
  \ref{mtl}(f) vanishes as well, but this is a consequence of considering the
  color-singlet state of the sources.
\item Two-loop vertex and gluon self-energy corrections, where the number
  of diagrams is also reduced by our choice of gauge, and double insertions
  of the corresponding one-loop corrections.
\item The graphs containing the four-gluon vertex with all ends attached to
  the sources vanish in Feynman gauge as well.
\end{itemize}
All other two-loop graphs are already accounted for by the exponentiation.

The relevant diagrams can be evaluated in momentum space without
encountering any special difficulties. Using the integration by parts method
\cite{IbP}, most of the integrals that occur can be reduced to products or
convolutions of the standard one-loop two-point function, its HQET equivalent
as given in \cite{BG91}, the HQET three-point function as given in
\cite{BBG93} and the mixed-type three-point function
$$
 \overline C({\bf p}^2) = \int\!\frac{d^Dl}{(2\pi\mu)^{-2\epsilon}}\Big(
   \frac{-1}{l^2\!+\!i\epsilon}\Big)^n\!\Big(\frac{-1}{(l\!+\!p)^2\!+\!
   i\varepsilon}\Big)^m\!\Big(\frac{-1}{lv\!+\!i\varepsilon}\Big)^a
$$
which can be computed by standard methods for $vp=0$. The diagrams
\ref{twoloop}(d--f), \ref{mtl}(e) and the vertex correction containing two
three-gluon vertices, however, require the computation of some true two-loop
integrals. As a detailed description of the calculation must be postponed to a
future publication, we only mention that the computer program FORM \cite{Form}
has been used for the evaluation of most of the diagrams and immediately
present the results.

Combining the two-loop result with the tree-level and one-loop expressions,
the effective coupling constant introduced above can be written as
\begin{eqnarray}\lefteqn{
  \alpha_{\rm V}({\bf q}^2)=\alpha_{\rm _{\overline{MS}}}({\bf q}^2)
    \Bigg\{ 1+\frac{\alpha_{\rm _{\overline{MS}}}({\bf q}^2)}{4\pi}\frac{31C_A
      -20T_Fn_f}{9}}&& \nonumber \\
  && +\Big(
     \frac{\alpha_{\rm _{\overline{MS}}}({\bf q}^2)}{4\pi}\Big)^2\Big[
     C_A^2 \Big( 
     \frac{4343}{162}+\frac{24\pi^2-\pi^4}{4}+\frac{22}{3}\zeta_3 \Big)
     \nonumber \\
  & & \quad
      -C_AT_Fn_f\Big( \frac{1798}{81}+\frac{56}{3}\zeta_3\Big) \nonumber \\
  & & \quad-C_FT_Fn_f\Big( \frac{55}{3} - 16\zeta_3 \Big) + 
      (T_Fn_f)^2\frac{400}{81}\Big] \Bigg\}. \label{result}
\end{eqnarray}
By inserting the three-loop running coupling in the $\overline{\rm MS}$-scheme
(the formula can be found, for example, in \cite{RPP}), we thus obtain the
three-loop potential, in the sense that the expression is correct up to a
constant multiplying $\alpha_{\rm _{\overline{MS}}}^3$. The terms proportional
to $C_FT_F$ and $T_F^2$ in (\ref{result}) could have been obtained from the
one- and two-loop gluon propagator, but the other two terms really required
computing.

Eq.\ (\ref{result}) can be used to determine the scheme-dependent
coefficient $\beta_2$ of the $\beta$-function for the $V$-scheme, as defined by
\begin{equation}
  \frac{1}{\alpha_{\rm V}}\frac{d\alpha_{\rm V}}{d\ln\mu^2} =
    -\sum_{n=0}^{\infty}\beta_n^{\rm (V)}\Big(\frac{\alpha_{\rm V}}{4\pi}\Big)
    ^{n+1},
\end{equation}
with the result (the first two coefficients, of course, coincide with those
of the $\overline{\rm MS}$-scheme)
\begin{eqnarray}
  \beta_2^{\rm (V)} & = & \Big(\frac{618+242\zeta_3}{9}+\frac{11(24\pi^2-\pi^4)
    }{12}\Big)C_A^3 \nonumber \\ &&
    -\Big(\frac{445+704\zeta_3}{9}+\frac{24\pi^2-\pi^4}{3}\Big)C_A^2T_F
    n_f \nonumber \\ &&
    + \frac{2+224\zeta_3}{9}C_A(T_Fn_f)^2 -\frac{686-528\zeta_3}{9}C_AC_FT_Fn_f
    \nonumber \\ &&
    +2C_F^2T_Fn_f+\frac{184-192\zeta_3}{9}C_F(T_Fn_f)^2.
\end{eqnarray}

The relation between the two couplings can, of course, be inverted easily,
yielding
\begin{eqnarray}\lefteqn{
  \alpha_{\rm _{\overline{\rm MS}}}({\bf q}^2)=\alpha_{\rm V}({\bf q}^2)
    \Bigg\{ 1-\frac{\alpha_{\rm V}({\bf q}^2)}{4\pi}\frac{31C_A-20T_Fn_f}{9}}
     && \nonumber \\
  && -\Big(\frac{\alpha_{\rm V}({\bf q}^2)}{4\pi}\Big)^2\Big[
     C_A^2 \Big(\frac{499}{162}+\frac{24\pi^2-\pi^4}{4}+\frac{22}{3}\zeta_3
     \Big)\nonumber \\
  & & \quad
      +C_AT_Fn_f\Big( \frac{682}{81}-\frac{56}{3}\zeta_3\Big) \nonumber \\
  & & \quad-C_FT_Fn_f\Big( \frac{55}{3} - 16\zeta_3 \Big) - 
      (T_Fn_f)^2\frac{400}{81}\Big] \Bigg\}.
\end{eqnarray}
This formula could be used to improve the precision when extracting $\alpha_{
\rm _{\overline{MS}}}(M_Z^2)$ from measurements of the Wilson loop on
the lattice \cite{MS96}.

For $N=3$ the results in actual numbers read
\begin{eqnarray}\lefteqn{
  \alpha_{\rm V}({\bf q}^2)=\alpha_{\rm _{\overline{MS}}}({\bf q}^2)
    \Bigg\{ 1+\frac{\alpha_{\rm _{\overline{MS}}}({\bf q}^2)}{\pi}\Big(2.583
    -0.278n_f\Big)}&& \nonumber \\
  && +\Big(\frac{\alpha_{\rm _{\overline{MS}}}({\bf q}^2)}{\pi}\Big)^2\Big(
    39.650 - 4.147 n_f + 0.077 n_f^2\Big)\Bigg\}, \\
  \lefteqn{\alpha_{\rm _{\overline{MS}}}({\bf q}^2)=\alpha_{\rm V}({\bf q}^2)
    \Bigg\{ 1-\frac{\alpha_{\rm V}({\bf q}^2)}{\pi}\Big(2.583
    -0.278n_f\Big)}&& \nonumber \\
  && -\Big(\frac{\alpha_{\rm V}({\bf q}^2)}{\pi}\Big)^2\Big(
     26.303 - 1.277 n_f - 0.077 n_f^2\Big)\Bigg\},
\end{eqnarray}
which shows that the coefficients of the second order terms are not small
even for $n_f=5$.

I would like to thank Prof. J. H. K\"uhn for suggesting this problem to me and
for carefully reading the manuscript. This work was supported by the
``Landesgraduiertenf\"orderung'' at the University of Karlsruhe.

\end{document}